%% file: ms_v2.tex
\documentclass[manuscript]{aastex}
\shorttitle{Sulfur in Orion}
\shortauthors{Daflon et al.}
\begin{document}

\title{Sulfur Abundances in the Orion Association B Stars} 

\author{Simone Daflon\altaffilmark{1}, Katia Cunha\altaffilmark{1,2}, 
Ramiro de  la Reza\altaffilmark{1}, Jon Holtzman\altaffilmark{3}, 
Cristina Chiappini\altaffilmark{4}}

\affil{$^1$ Observat\'orio Nacional, Rua General Jos\'e Cristino 77 \\
 CEP 20921-400, Rio de Janeiro  Brazil }
\affil{$^2$ National Optical Astronomy Observatories, 950 N. Cherry Ave., 
Tucson, AZ 85719 }
\affil{$^3$ New Mexico State University, 1320 Frenger St.\\
Las Cruces, NM 88003}
\affil{$^4$ Observatoire de Gen\`eve, Universit\'e de Gen\`eve, 51 Chemin des Maillettes, 
CH-1290 Sauverny, Switzerland}
\email{daflon@on.br}

\clearpage
\begin{abstract}
Sulfur abundances are derived for a sample of ten B main-sequence 
star members of the Orion association. 
The analysis is based on LTE plane-parallel model atmospheres and 
non-LTE line formation theory by means of a self-consistent
spectrum synthesis analysis of lines from two ionization states of sulfur, 
S~{\sc ii} and S~{\sc iii}. The observations are high-resolution
spectra obtained with the ARCES spectrograph at the Apache Point
Observatory. The abundance distribution obtained for the Orion targets
is homogeneous within the expected errors in the analysis: A(S)=7.15$\pm$0.05. 
This average abundance result is in agreement with 
the recommended solar value (both from modelling of the photospheres  
in 1-D and 3-D, and meteorites) and indicates that little, 
if any, chemical evolution of sulfur has taken place in the last
$\sim$4.5 billion years. 
The sulfur abundances of the young stars in Orion are found to agree well
with results for the Orion nebulae, and place strong constraints on the amount 
of sulfur depletion onto grains as being very modest or nonexistent.
The sulfur abundances for Orion are consistent with other measurements at a similar 
galactocentric radius: combined with previous results for
other OB-type stars produce a relatively shallow sulfur abundance gradient
with a slope of $-0.037 \pm$0.012 dex Kpc$^{-1}$.
\end{abstract}

\keywords{open clusters and associations: individual (Ori OB1) --- 
stars: abundances --- stars: early-type}

\newpage

\section{Introduction}\label{int}

The chemical element sulfur, together with O, Ne, Mg, Si, Ar and Ca, belongs to the 
family of the so-called $\alpha$-capture elements, which are light elements with even Z in 
the range between 6 $<$ Z $<$ 22. Sulfur is among the ten most abundant 
elements in the universe and its production occurs both during hydrostatic and 
explosive oxygen-burning phases in massive star evolution and supernovae of Type {\sc ii} 
(SN~{\sc ii}; Clayton 2003). Although some contribution from SN {\sc i}a's is also 
expected (Fran\c cois 2004, and references therein). 

The observed evolution of the sulfur abundance in the Galaxy is generally less well-defined
than that of other $\alpha$-elements (such as oxygen or calcium). The earliest determinations 
of sulfur abundances in stars dates from 1980's \citep{ctl81,fra87,fra88}. More recently, 
\citet{che02} studied S~{\sc i} lines in a sample of disk stars (with [Fe/H] $>$ -1.0) 
and found that sulfur abundances correlate with those of silicon, as
generally expected since these are both considered to be $\alpha$-elements. 
Concerning the behavior of [S/Fe] versus [Fe/H] in low-metallicity halo stars, \citet{ier01}
and \citet{tak02} found that [S/Fe] increased linearly with decreasing [Fe/H], while
\citet{rel04} and \citet{nis04} obtained values of [S/Fe] that are roughly constant 
and enhanced relative to solar value for metal poor stars, following the general trend 
observed for other $\alpha$-capture elements. \citet{caf05}, on the other hand, found at 
low metallicities both a population of stars having a flat [S/Fe] $\sim$ +0.4 dex, and 
stars with higher [S/Fe] ratios. The recent results by \citet{nis07}, which take into account 
corrections for non-LTE effects, indicate that halo stars follow a plateau at [S/Fe] $\sim$ +0.2 dex.

An important property of sulfur, a volatile element, is that given its low condensation temperature
it is not expected to locked-up significantly into grains, at least not in the low-density
interstellar medium, nor in certain nebular sites \citep{goi06}. Thus, a
comparison between nebular and stellar sulfur abundances may involve relatively small
corrections for grain depletion; if true, sulfur then would provide
a direct connection between the stellar and gas-phase abundances in 
the interstellar medium, as well as nebular abundances in galactic and extra-galactic environments.
This work  will focus on defining the present-day stellar sulfur abundance in 
a sample of young B-star members of the Orion association.
These abundances can be used to help define the present-day sulfur abundance in the solar
neighborhood, as well as constrain sulfur depletions onto grains. In addition,
the Orion sample provides another data point to define the Galactic disk sulfur
abundance gradient.

\section{Observational Data}\label{obs}

\begin{figure}
\epsscale{1.0}
\plotone{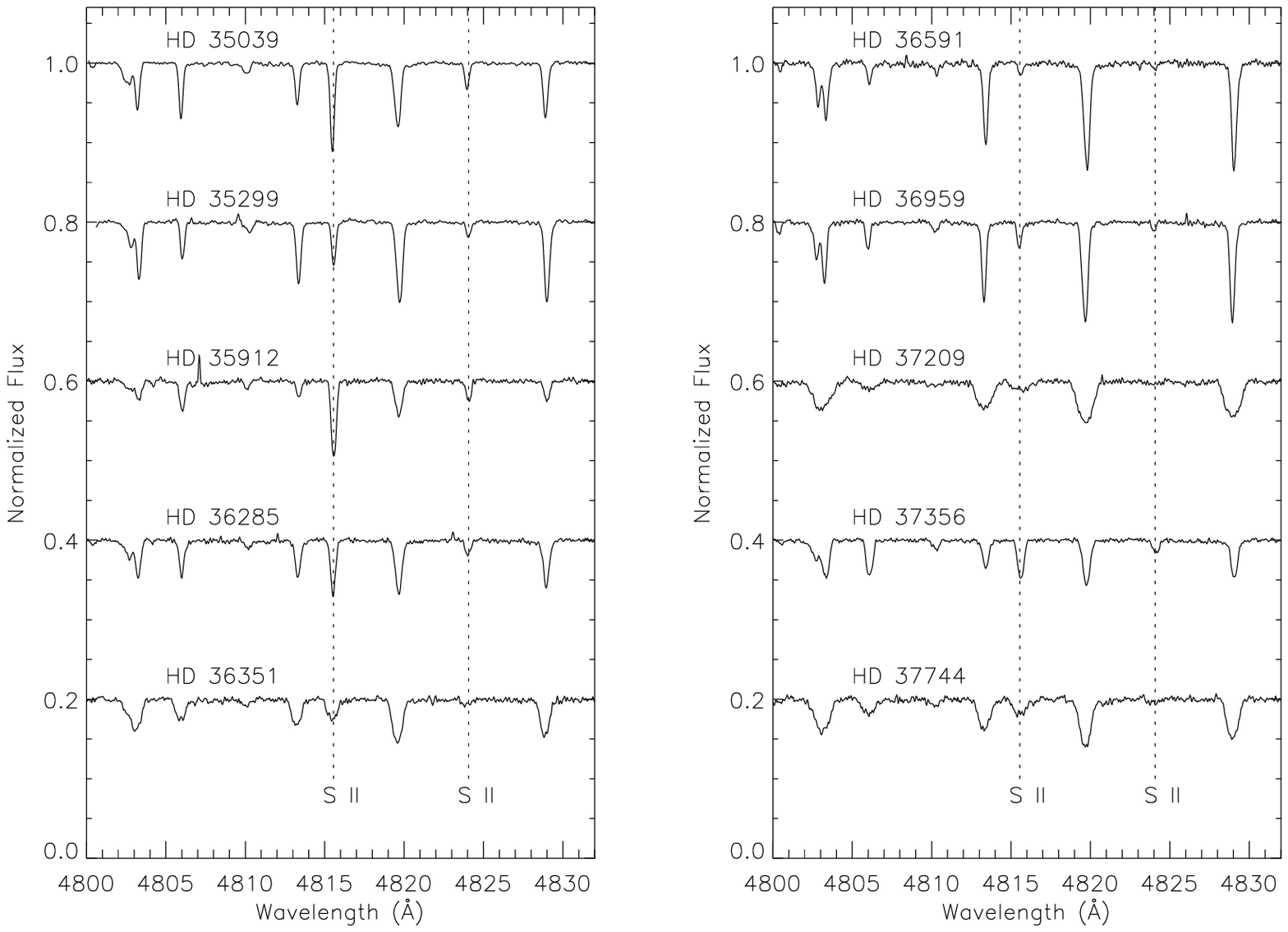}
\caption{Sample observed spectra in the spectral region between 4800--4832\AA \ for all 
observed targets.
\label{fig1}}
\end{figure}

The targets are 10 main-sequence early B-type star members of the Ori OB1 association, 
selected from the sample of \citet{cel94} and previously analyzed for neon abundances
in \citet{chl06}.
The observational data are high resolution (R$\sim$35,000) spectra obtained with 
the  Astrophysical Research Consortium (ARC) Echelle Spectrograph on the 3.5m telescope 
at the Apache Point Observatory in 2007. 
The spectra cover the wavelength range between 3480--10260\AA \ and have signal-to-noise ratios 
generally greater than $\sim$100. The data were reduced and the spectra were extracted and 
normalized to a unity continuum  following standard procedures. In Figure~\ref{fig1} we show 
sample spectra for all target stars in the spectral region  between 4800\AA--4832\AA \, 
where the S~{\sc ii} lines at 4815\AA\ and 4824\AA\ are found.
More details about the observations and data reduction can be found in \citet{lan08}.

\section{Stellar Parameters and Sulfur Abundance Determinations}\label{ana}

The stellar parameters, effective temperature, surface gravity, and
microturbulent velocity, adopted in this analysis are listed in
Table~\ref{tbl-1} and taken from \citet{cel94}. The values of 
T$_{\rm eff}$ and $\log g$ in that study were
obtained from an iterative scheme which combines photometric calibrations for the 
Str\"omgren indices $c_o$, [$c_1$], and $\beta$ and fitting of LTE theoretical 
profiles to the H$\gamma$ line wings.  
The microturbulence, $\xi$, was derived from an
analysis of O~{\sc ii} lines, by requiring that the oxygen abundances are
independent of the line strength. 

Sulfur abundances for a sample of 16 S~{\sc ii} and 3 S~{\sc iii} lines 
were derived from non-LTE synthetic profiles and adopting LTE model atmospheres 
from \citet{kur93} plus a sulfur model atom from  \citet{vra96}. 
Our line sample was selected from the Kurucz linelist and included all S~{\sc ii} and 
S~{\sc iii} lines within 4000 -- 5100 \AA\ with profiles which could be
synthesized (corresponding to a minimum equivalent width $\sim$ 2 m\AA) and which were 
relatively free of blends.
The adopted model atom treats S~{\sc ii}/S~{\sc iii} simultaneously with  
81 levels of S~{\sc ii} and 21 levels of S~{\sc iii} and also includes the 
three lowest levels of S~{\sc i} and the two lowest levels of S~{\sc iv}, together 
with the ground state of S~{\sc v}.    
The solutions to the statistical equilibrium  and transfer equations were
obtained with the program DETAIL \citep{gid81} which assumes LS-coupling.

\begin{figure}
\epsscale{1.0}
\plotone{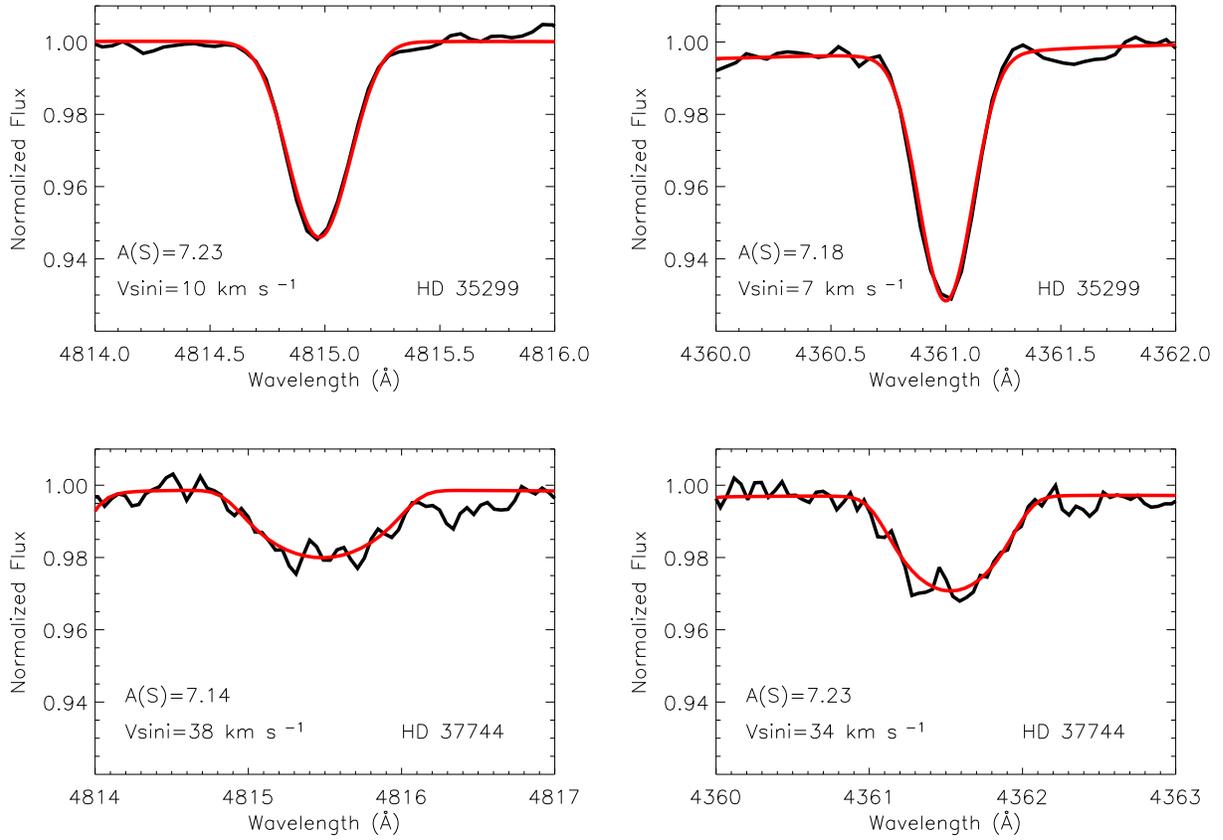}
\caption{Sample fits showing synthetic profiles (red line) and observed spectra (black line) 
of a S~{\sc ii} line at 4815 \AA\ and a S~{\sc iii} line at 4361 \AA\ for the target stars
HD~35299 and HD~37744. The best-fit $V \sin i$ and abundance values for each line are shown 
in the panels.
\label{fig2}}
\end{figure}

Synthetic line profiles were calculated with the code SURFACE \citep{beg85} and assuming 
Voigt profile functions. These profiles were then broadened by means of 
convolution with a rotational profile, including both the projected rotational
velocity ($V \sin i $), limb 
darkening, and the instrumental profile. The  microturbulence for each  
star was kept constant while abundances and $V \sin i$'s were allowed to vary. 
The best fit for each line was obtained from the $\chi^2$-minimization of the differences 
between theoretical and observed profiles. 
The S~{\sc ii} and S~{\sc iii} transitions analyzed and atomic data (excitation potentials 
and oscillator strengths) are found in Table~\ref{tbl-2}; the $gf$-values are from the 
Opacity Project \citep{cem92}. In Figure~\ref{fig2} we show examples of best-fit synthetic 
profiles of S~{\sc ii} and S~{\sc iii} lines for the target stars HD~35299 and HD~37744.

\begin{figure}
\epsscale{1.0}
\plotone{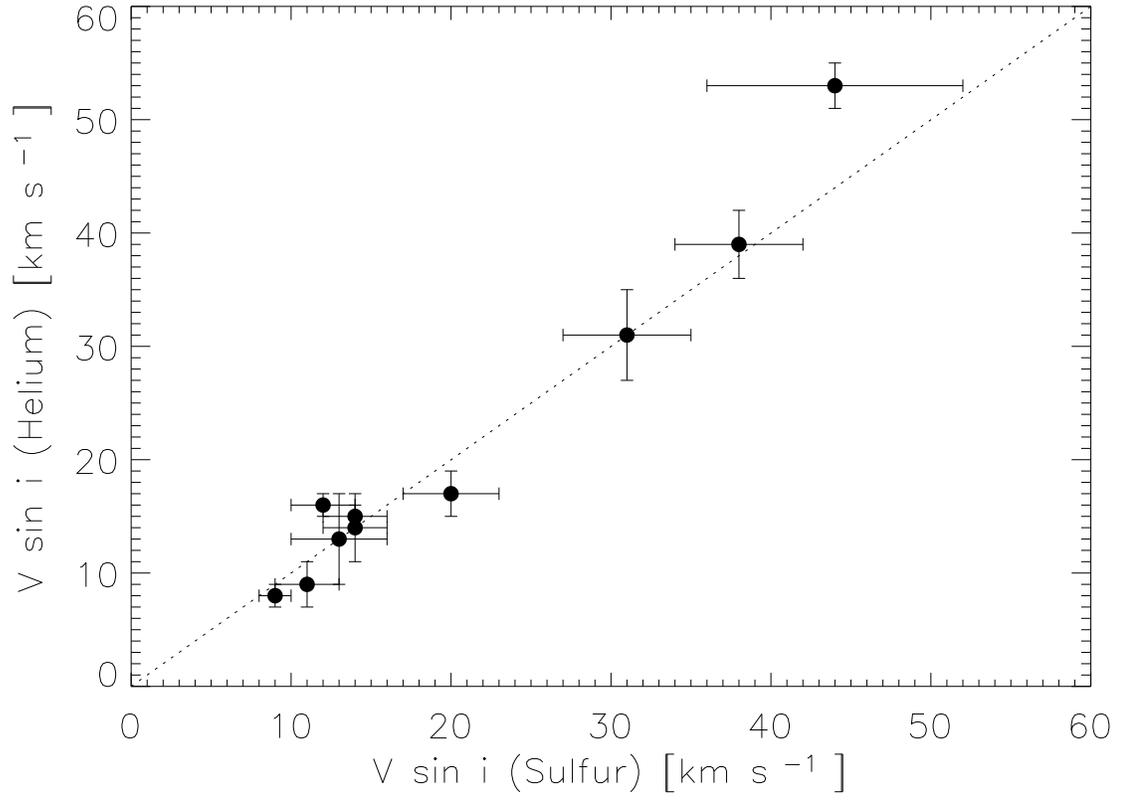}
\caption{The comparison between projected rotational velocities derived from the fitting of sulfur 
lines (this study) with $V \sin i$ results based on a calibration of FWHM of theoretical 
non-LTE He~{\sc i} profiles.  
\label{fig3}}
\end{figure}

Final sulfur abundance results (including S~{\sc ii} and  S~{\sc iii} abundances) 
and $V \sin i$ values for the stars are presented in Table~\ref{tbl-1}: these are the averages for
the individual lines and the respective abundance dispersions are also listed
(with the number of sulfur lines fitted for each star appearing in brackets). 
We note that the dispersions presented in the table represent the line-to-line 
scatter and are not representative of real uncertainties in the derived abundances. 
The total errors in our abundance analysis are estimated in Section~\ref{err}. 
As a check, we compared the projected rotational velocities in Table~\ref{tbl-1} 
with $V \sin i$'s derived by adopting a calibration by \citet{daf07} of full-widths at half
maximum of He~{\sc i} lines at $\lambda\lambda$4026, 4388 and 4471\AA\ 
measured from a grid of synthetic spectra calculated in non-LTE.
The comparison between the $V \sin i$'s obtained from sulfur lines
with those using the He lines  calibration is shown in Figure~\ref{fig3}. We note a very good 
agreement between the two sets of $V \sin i$ values; except one star 
(HD~37209) for which the two determinations agree marginally, within the uncertainties  
($V \sin i_{Sulfur} = 44 \pm 8$ km s$^{-1}$ and $V \sin i_{Helium} = 53 \pm 2 $km s$^{-1}$). 

\begin{figure}
\epsscale{1.0}
\plotone{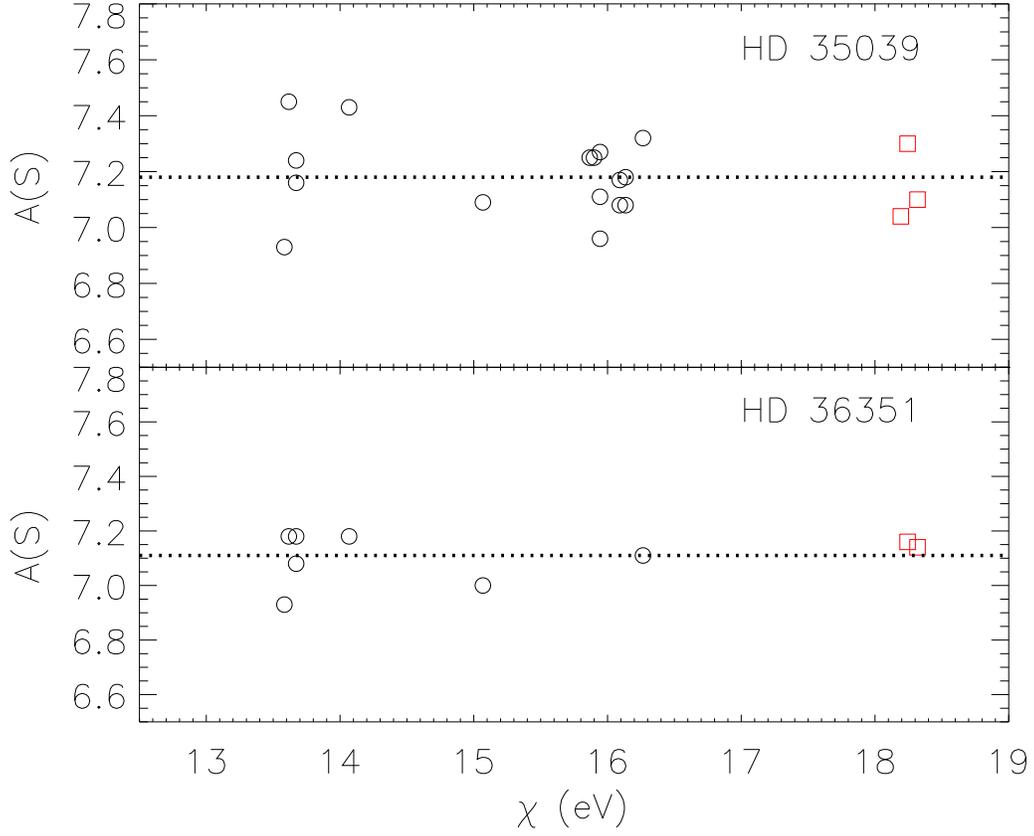}
\caption{Sulfur abundances of individual S~{\sc ii}  (open circles) and
S~{\sc iii}  (open squares) lines as a function of the ionization potential $\chi$, showing the
sulfur ionization balance for the stars HD~35039 (top panel)  and HD~36351 (bottom panel).
The effective temperatures are from Table 1. The dotted lines represent the mean sulfur abundance 
for each star.
\label{fig4}}
\end{figure}

The analysis of sulfur in B stars is aided by the fact that lines from two ionization stages
(S~{\sc ii} and S~{\sc iii}) can be observed and this offers the possibility to constrain the 
effective temperatures from the ionization balance between S~{\sc ii} and S~{\sc iii}. 
Spectroscopic T$_{\rm eff}$s, obtained from an agreement between the abundances from 
S~{\sc ii} and S~{\sc iii} lines, can provide an additional  check on the adopted effective 
temperature scale for the studied stars (shown in Figure~\ref{fig4}). For most stars in our 
sample, sulfur abundances derived from  S~{\sc ii} and S~{\sc iii} lines (adopting the
T$_{\rm eff}$-scale from Cunha \& Lambert 1994) were found to agree within $<$ 0.10 dex.
We note however, that the stars HD~36351, HD~36591, and HD~37744 yielded differences in 
A(S~{\sc iii} $-$ S~{\sc ii}) of +0.20 dex, +0.16 dex and $-0.17$ dex, respectively.
The strength of the S~{\sc ii} lines reaches a maximum around T$_{\rm eff}$=18,000--19,000 K,
whereas the S~{\sc iii} lines are more sensitive to  T$_{\rm eff}$ variations in this 
temperature range (See figures 7 - 10 of Vrancken et al. 1996).  
It is likely that the differences between the abundances derived from 
S~{\sc ii} and S~{\sc iii} lines result from uncertainties in the adopted temperatures.
We thus revised the effective temperatures of these three stars by +2.6\%, +2\% and $-$2\%, 
respectively, in order to bring S~{\sc ii} and S~{\sc iii} abundances into agreement. 
The $\log g$ values for these stars were also revised in order to obtain similar theoretical fits 
to the H$_\gamma$ profiles.
These revised T$_{\rm eff}$ and $\log g$ values and the respective abundance results are 
listed in Table~\ref{tbl-1}.  

\begin{figure}
\epsscale{1.0}
\plotone{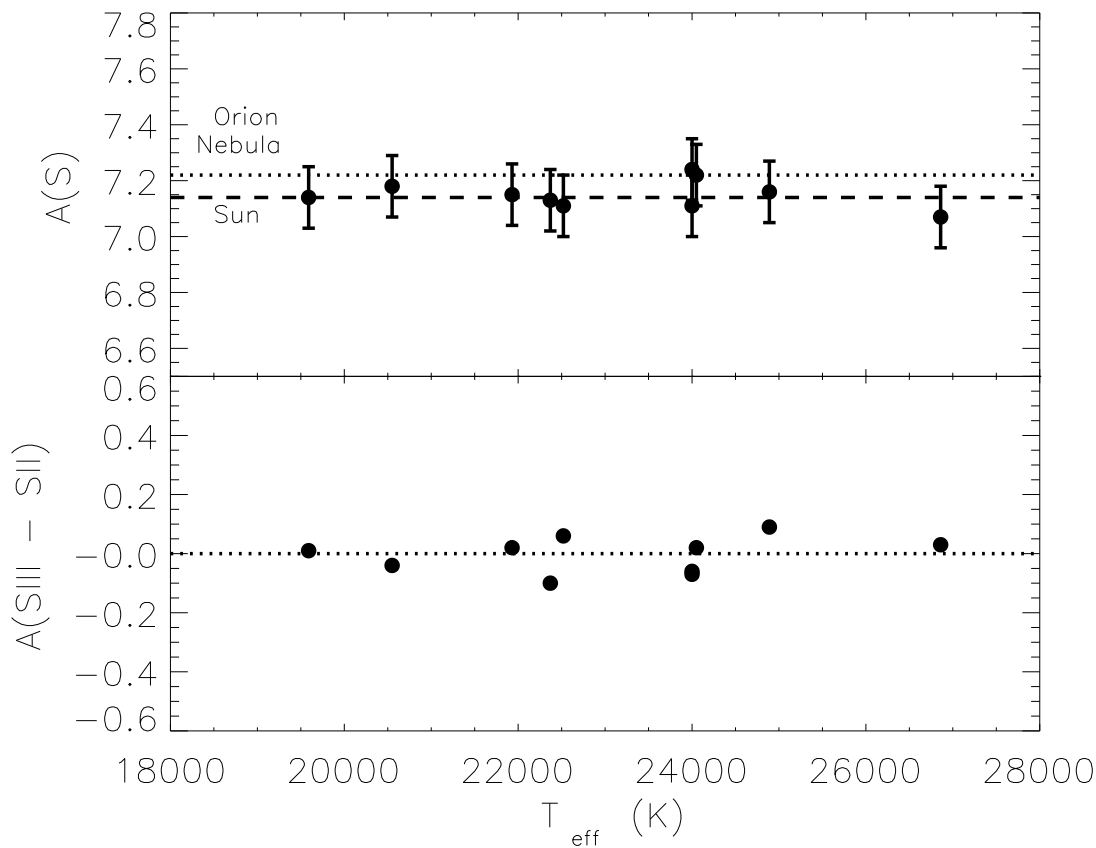}
\caption{Top panel: Sulfur abundance results for the sample stars as a function of
the adopted effective temperatures. The individual stellar values (filled circles)
are the mean sulfur abundances which were obtained from the measured lines and the error bars
represent the total errors from Table 3. We also show the abundance measured for
the Orion Nebulae by Esteban et al. (2004; dotted line) and the recommended
solar value by Asplund et al. (2006; dashed line).
Bottom panel: The differences between sulfur abundances derived
from S~{\sc iii} and S~{\sc ii} lines as a function of the adopted effective
temperatures.
\label{fig5}}
\end{figure}

Sulfur abundance results for the Orion stars are plotted as a function of the adopted effective 
temperature in the upper panel of the Figure~\ref{fig5}. The effective temperatures of the 
stars in this sample encompass the range roughly between 20,000K--27,000K and, although for 
B stars this represents a relatively broad range in T$_{\rm eff}$, no trends with abundance 
are observed. The flat distribution of sulfur abundances derived suggests that these results 
are probably free of major systematics within this T$_{\rm eff}$ range. In the lower panel of
Figure~\ref{fig5}, we present the difference between the abundances derived from S~{\sc iii} 
and S~{\sc ii} lines (A(S~{\sc iii} $-$ S~{\sc ii})) also plotted as  a function of the adopted 
T$_{\rm eff}$. The differences are small, $<$ $\sim$0.10 dex, and do not show
a significant trend with the effective temperature. 

In order to briefly evaluate the importance of adopting a treatment in non-LTE versus LTE when
studying sulfur abundances in B-type stars, we computed LTE abundances for one
target star HD~35299 with T$_{\rm eff}$=24,000 and $\log g$=4.25. The results from this test 
calculation indicate a modest correction for the average abundance of S~{\sc ii} lines: 
$\delta$ (S~{\sc ii})$_{NLTE} -$ (S~{\sc ii})$_{LTE} = -0.03 $dex. 
This is a small effect in the average abundance but the S~{\sc ii}  line-to-line scatter 
in LTE is found to be much larger ($\sim$ 0.25 dex) than in non-LTE ($\sim$ 0.15 dex). 
It should be noted that some of the S~{\sc ii} lines, such as $\lambda$ 4278.5\AA, have 
significant non-LTE corrections ($\sim  -$0.7 dex); while others, such as $\lambda$ 
4269.7\AA\ and $\lambda$ 4294.4 \AA\, have relatively small non-LTE corrections 
($\sim$ +0.04 dex).  A reduced S~{\sc ii} scatter in the average non-LTE abundance as 
opposed to LTE generally indicates that there is good consistency in the non-LTE calculations 
of individual lines and adequacy of the adopted model atom. 
The non-LTE corrections for the S~{\sc iii} lines studied here are also not significant: 
$\delta$ (S~{\sc iii})$_{NLTE} -$ (S~{\sc iii})$_{LTE} = -0.02 $dex; 
the line-to-line S~{\sc iii} scatter both in LTE and non-LTE are found to be quite small. 

\subsection{Abundance Uncertainties}\label{err}

The abundance results in this study are based on a spectrum synthesis analysis and
are subject to uncertainties arising mainly from the uncertainties in stellar parameters, 
microturbulent velocities, as well as placement of the continuum and gf-values. 
In order to estimate the errors in the derived sulfur abundances, 
we re-computed the abundances for HD~35299 (a sharp-lined star with T$_{\rm eff}\sim$24,000K)
by independently increasing each of the following parameters, one at a time, by:
4\% for  T$_{\rm eff}$; 0.1 dex for $\log g$; 1.5 km s$^{-1}$ for microturbulence; 
0.5\% for the continuum location and 10\% for gf-values.
The sensitivity of the sulfur abundances to variations in these parameters are
presented in Table~\ref{tbl-3}.  
The combined errors in the derived S~{\sc ii} and S~{\sc iii} abundances are
0.11 dex and 0.10 dex, respectively. 

The stellar parameters in this study were taken from Cunha \& Lambert (1994) and these
were obtained from an approach which combined Str\"omgren photometry and LTE theoretical profile fits 
to the H$_\gamma$ line wings (Section 3). The adoption of LTE model profiles likely overestimates 
$\log g$s, in particular for hotter stars with T$_{\rm eff} \sim$30,000K \citep{nep07}. 
For the effective temperature range of the target stars, however, this effect is probably within 
the uncertainties in the determinations. In order to quantify the errors in log g we did test 
calculations by re-deriving $\log g$s for 2 target stars (the coolest and the hottest in our sample) 
using NLTE thoretical profiles of H$_\gamma$ calculated with DETAIL and SURFACE (Butler 2000, private
communication). 
The results obtained indicate that the differences between ($\log g_{\rm LTE}$ - $\log g_{\rm NLTE}$) 
vary between 0.05 dex (for the coolest star in our sample) and 0.1 dex (for the hottest star in our 
sample). Such differences in $\log g$ would result in sulfur abundance differences between $-$0.01 and 
$-$0.02 dex, at most (Table 3). 

\section{Discussion}\label{dis}

Non-LTE sulfur abundances are derived here for a sample of young B stars from the Orion
association: the abundance distribution obtained in this study is homogeneous
with a relatively small scatter which is roughly of the order of the abundance errors.
The average sulfur abundance for our sample is A(S)= 7.15$\pm$0.05. In the following
section we will compare these results with the solar abundance.

\subsection{Sulfur Abundances in Orion and the Solar Value}
 
In recent years, the sulfur abundance which is recommended for the Sun has been 
revised downward by $\sim$0.15--0.20 dex (see, for example, Lodders et al. 2009). 
The compilation of Solar System abundances by \citet{ges98} listed
a photospheric abundance of A(S)= 7.33$\pm$0.11. More recently, \citet{ags06}
did a critical evaluation of sample S~{\sc i} lines and removed those lines which were 
deemed to be blended. This study recommended a lower sulfur abundance of A(S)=7.14 $\pm$ 0.05 
for the solar photosphere, in much better agreement with measurements in meteorites
(A(S)=7.17 $\pm$ 0.02, Lodders et al. 2009).
Note that the photospheric abundances discussed above were obtained using 1-D model 
atmospheres. An important confirmation of the 1-D result comes from the recent modelling
of the forbidden [S~{\sc i}] line at 1082nm by  \citet{cel07}, which is based on 3-D 
hydrodynamical model atmospheres, that finds a sulfur abundance extremely consistent 
with the 1-D results: A(S)= 7.15$\pm(0.01)_{stat}\pm(0.05)_{sys}$.

The average sulfur abundance of the Orion B stars obtained here is found to be
in perfect agreement with the Solar System value: the abundances all cluster around A(S)=7.15. 
Such an agreement, taken at face value, indicates that a non-measurable evolution of 
the sulfur abundances has taken place in the last $\sim$4.5 Gyrs. The consistency between 
the sulfur abundances measured for Orion and the Sun (both from 1-D and 3-D model atmospheres 
and meteorites) also argues favorably for the recent claim by \citet{chl06}
and \citet{lan08} that the abundances of Ne and Ar measured in the Orion association can
be considered as good proxies for the Solar System abundances; this connection is especially 
interesting in the case of noble gases because the solar abundances of such elements are much more 
uncertain, yet these abundances need to be defined as they affect solar interior models.

\subsection{Previous Results for Early-Type Stars}

Recent studies of chemical abundances
in OB stars have not analyzed sulfur (e.g. Pryzbilla et al. 2008; Simon-Diaz et al. 2006)
and most of the sulfur abundances which are available in the literature are based on 
LTE treatments. For example, \citet{gel92} presented a detailed LTE analysis for a 
sample of early-type B stars focusing on CNO,  but also derived
S~{\sc ii} abundances for the coolest B stars in their sample (with  T$_{\rm eff}$ $<$ 24,750K).
The mean S abundance in that study is A(S)= $7.21\pm0.12$, with a slight increase in the 
abundance as a function of effective temperature that could be explained in terms of the expected
errors.  On the other hand,   \citet{kil94} analyzed 
a sample of seven stars in the Orion association and obtained $<A(S)>_{LTE} = 6.80\pm0.25$. 
Their derived abundances show a trend with the effective temperature, and the authors 
suggest this may result from neglecting non-LTE effects in their analysis.  

Line formation in non-LTE was considered in the study by \cite{vra96}.
The focus in that paper was on the construction and description of a fairly complete 
sulfur model atom (which is adopted in the present analysis), but the authors also derived
abundances for three B stars as a test of their model atom, for which they obtained sulfur 
abundances in  the range A(S)=7.14 -- 7.38. 
One of the stars studied in that paper, in particular, is also in our sample, HD~37356.
For this star, they obtained A(S) = $7.14\pm0.16$ (considering only 
S~{\sc iii} lines), in full agreement with A(S) = $7.13\pm0.14$, 
obtained here from 13 S~{\sc ii} and 3 S~{\sc iii} lines. 
More recently, \citet{mor06} used the same methodology as in  \cite{vra96}
and analyzed sulfur abundances (along with other elements) in a sample of nine $\beta$ 
Cephei stars. The average sulfur abundances found was A(S)=7.21 $\pm$ 0.13.
The results from this sample was later combined with sulfur abundances
of 11 stars (not all $\beta$ Cephei; Morel et al. 2008) and the average sulfur abundance 
obtained is A(S)=7.18 $\pm$ 0.11. We note, however, that the results obtained
for Cepheid stars  \citep{and02} are significantly higher than the above:
the average sulfur abundance calculated for a sample of 24 Cepheids in the solar
neighborhood (located within 7.6 and 8.2 kpc from the Galactic center) is A(S)=7.44 $\pm$ 0.09. 

The first systematic non-LTE abundance analysis of sulfur in a large sample of OB stars 
is presented in the series of papers by \citet{daf03,daf04a, daf04b}. 
The sulfur abundances in these previous studies from our team, however, were based only on two
weak S~{\sc iii} lines at 4361 and 4364\AA, which fall close to the edge of the red wing
of $H_\gamma$. It should be noted that for the majority of their targets, 
the S~{\sc iii} line at 4364\AA\ was the $\it{only}$ measured line, even for target stars 
with the lowest $V \sin i$s and having spectra with the highest signal-to-noise ratios. 
The target sample analyzed in these studies covered mostly the solar neighborhood and the 
inner Galactic disk (galactocentric distances $R_g$ between $\sim$ 5--9 kpc) but also included 
4 more distant targets  ($R_g \sim$ 9--14 kpc) which helped in the definition of disk 
metallicity gradients as discussed in \citet{dec04}.

\subsubsection{Sulfur Abundance Gradients}

\begin{figure}
\epsscale{1.0}
\plotone{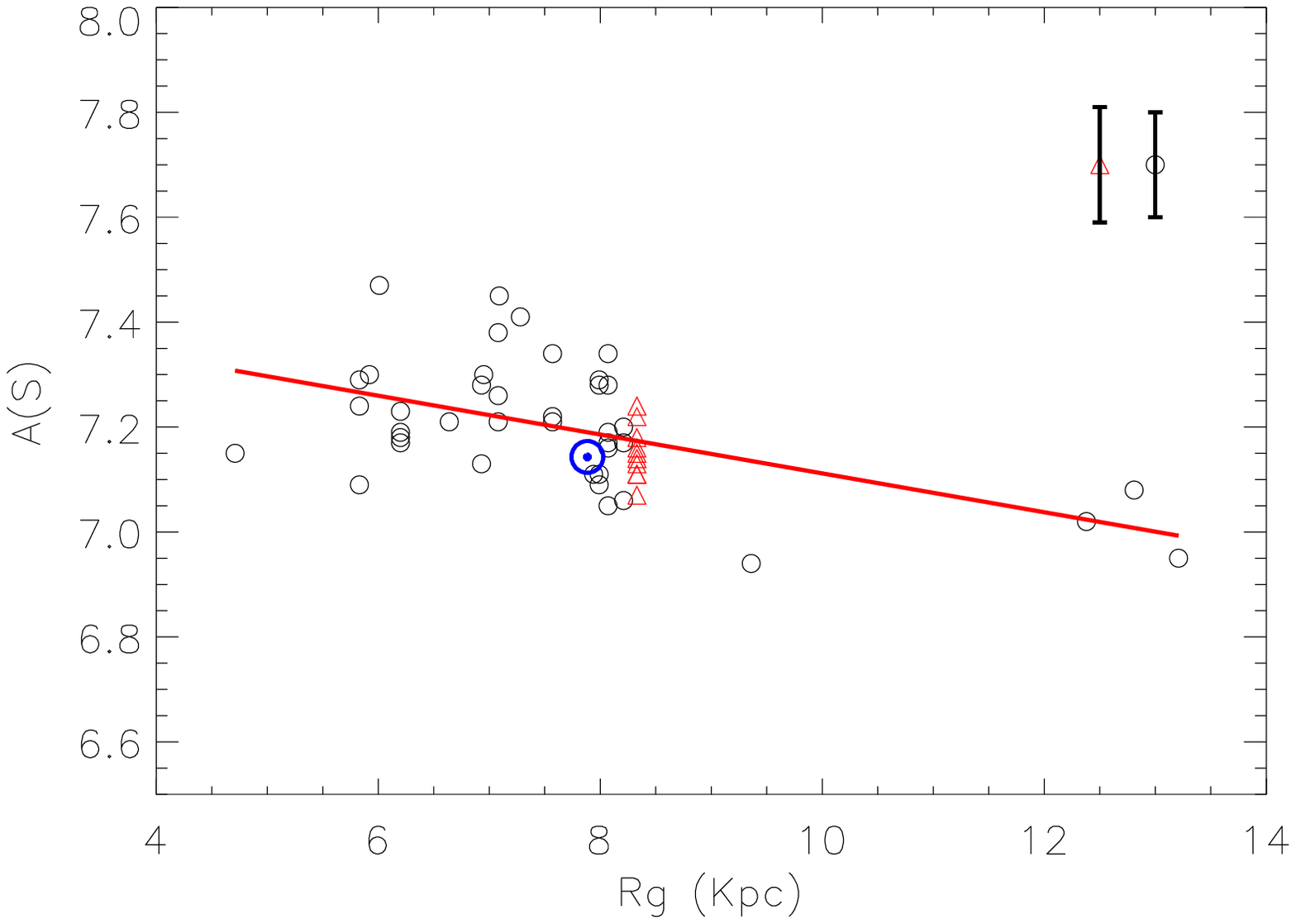}
\caption{ Sulfur abundances plotted as a function of galactocentric distances. The black open circles
represent the sample of OB stars studied by \citet{dec04}; the red open triangles
are the Orion abundances for the B stars analyzed in this paper. The Sun is represented at 
R$_g$=7.9 Kpc \citep{mac00} with $A(S)=7.14$ from \citet{ags06}. 
The sulfur gradient calculated for the entire sample has a slope of $-0.037 \pm$0.012 dex Kpc$^{-1}$ 
(shown as the solid red line). The typical errorbars in the abundances are shown in the top right.
\label{fig6}}
\end{figure}

The sulfur abundance results from \citet{dec04} are shown versus 
the adopted galactocentric distances in Figure~\ref{fig6} (open circles). 
The sulfur results for Orion are shown as open triangles 
and these seem to agree well with the overall trend.
The overlap between the abundance results in the solar neighborhood is clear.
In particular, if one averages the abundances of all stellar members of the five OB associations  
located within R$_g$=7.6--8.2 Kpc, which are Cep OB2, Cep OB3, Cyg OB3, Cyg OB7 and Lac OB1,
the average sulfur abundance for this nearby sub-sample is A(S)=$7.19\pm0.09$: this compares
well with the average sulfur abundance obtained for Orion, which is located $\sim$ 500 pc 
from the Sun,  and gives support to previous results \citep{pry08} that the solar neighborhood 
abundance is homogeneous, at least at the level of the uncertainties in the abundance analysis. 
 
Such agreement in the abundance results for the solar neighborhood is pleasing since the 
sulfur abundances derived here are based on several lines from two ionization stages 
(S~{\sc ii} and S~{\sc iii}), while the previous analyses were obtained from one, or a 
maximum of two, sulfur lines from a single ionization stage (S~{\sc iii}), and from a different
T$_{\rm eff}$-scale: the T$_{\rm eff}$s from \citet{dec04} are based on 
a calibration for the reddening-free parameter Q, and the effective temperatures from \citet{cel94}
are based on an iteractive scheme that combines photometric calibrations for 
Str\"omgren indices and the fitting of H$_\gamma$ wings.

Since there seems to be no major systematic differences between the sulfur abundances
previously published by our group and from this study, the Orion results can be added to compute 
a new sulfur gradient for the Galactic disk.
A linear fit to the run of abundances with distances from the Galactic center is a possible 
simple description for the overall decrease in the sulfur abundances towards the
edge of the disk. The best-fit straight line to the complete dataset in Figure~\ref{fig6} 
corresponds to $A(S)=(7.481\pm 0.097) -(0.037\pm 0.012) R_g$. The adopted distances are
from \citet{dec04}. The sulfur gradient obtained is flat, in line with the
previous result obtained by \citet{dec04}, and
very similar to the previously derived oxygen gradient for the same target sample.
A similar behavior between sulfur and oxygen is expected as these are both $\alpha$ elements.

\subsection{Sulfur Depletion}\label{gas}

The Orion Nebula is largely considered as a standard reference for the ionized gas abundances
in the local neighborhood (see, for instance, 
Esteban et al., 2004). However, the direct comparison between the abundance derived 
for a particular element in the Nebula and in the Sun should consider two important points: 
first, its depletion onto grains, and second, the effects of the chemical evolution of the solar
neighborhood since the Sun was formed. 
Concerning the evolutionary effects, chemical evolution models by \citet{crm03}
indicate only a small effect, of the order of $\sim$ 0.1 dex, on the sulfur evolution between  the
formation of the Solar System and the present.

Concerning the more general discussion about the depletion of sulfur onto grains, 
there are clear indications that sulfur is undepleted in diffuse interstellar gas. 
\citet{hss06} derive a gas-phase sulfur abundance of A(S) = $7.13 \pm0.03$  for the  
Galactic Halo insterstellar medium. In denser regions, however, the amount of sulfur 
depletion is not completely
clear. \citet{tie94} considered gas-phase abundances in S-bearing molecules and argued that
sulfur depletions of factors of $\sim 10^3$ could exist. Also, 
dominant $S^+$ ions (of this easily ionized element) can efficiently freeze-out 
onto negatively charged dust grains in denser regions \citep{ruf99}.
Puzzling contrary indications exist, as evidenced by the presence of S~{\sc ii}
recombination lines in dense clouds and the absence of strong infrared sulfur features in grains
-- See \citet{goi06} for a discussion.
One way to gain insight into this problem is to study Photodissociated Regions
(PDR), which are intermediate between diffuse and dark clouds
as, for example, the Horsehead PDR \citep{goi06}. Even in 
an environment with large chemical complexity,  low sulfur depletion has been found. 

In order to constrain the amount of sulfur depletion in the Orion H~{\sc ii} region, 
one can directly compare the sulfur abundances of young early-type star members of the 
OB association with the chemical composition of the Orion nebulae, albeit keeping in mind that 
there may be systematic abundance errors which  afflict both the stellar and nebular analysis 
independently.  The chemical composition of the gas which formed the young stars in the 
Orion association (and which can be measured in the stellar photospheres of OB stars) 
represents an independent check on the present day gas content, which is measured
from nebular lines in its H~{\sc ii} region.

The most recent and detailed study 
of the chemical composition of the Orion nebulae by \citet{est04} observed a region near the 
hot star $\theta^1$ Ori, and measured A(S)= $7.22\pm0.04$; when adopting their preferred value 
for the temperature fluctuation parameter, $t^2$ =0.02
(see Figure~\ref{fig5} showing the results obtained for the young stars in Orion comparison 
with the dotted line which represents the sulfur abundance for the Orion nebulae from 
Esteban et al. 2004). Although it should be kept in mind that the stellar and nebular abundances 
in Orion agree within the quoted errors in the determinations, a brief comparison of the 
results is of interest.
The difference between the nebular and stellar abundances is found to be positive 
($\delta$A (S)$_{Neb-Stel}$ = +0.07 dex) and not explainable as the
result of the chemical evolution sulfur as the Orion stars are very young, with ages 
less than $\sim$ 10 million years. 
Since the net effect of entrapment of gas onto grains is to reduce the gas phase abundance,
the nebular abundance could be viewed as a lower-limit abundance value.
If the lowest sulfur abundance value in \citet{est04} of A(S)=7.06 is adopted, 
(which corresponds to the extreme case of t$^{2}$=0.0; not favored by the authors),
the amount of sulfur depletion using the B stars as a reference 
would still be modest or non-existent within the uncertainties.
It is easily arguable from these results and their respective uncertainties, that
sulfur cannot be significantly depleted in the Orion nebulae. 

\section{Conclusions}\label{con}

Sulfur abundances are derived in non-LTE for a sample of young B stars of the Orion association.
The conclusions from this study are the following:

\begin{enumerate}
\item The abundance distribution in Orion is found to be homogeneous within the uncertainties
in the analysis.  The comparison of sulfur abundances derived for Orion B stars with abundance 
results for the Sun, the insterstellar medium and early-type stars in the solar vicinity allows 
us to conclude that  the sulfur abundances have undergone very little change from the time of 
formation of the Solar System $\sim$ 4.5 Gyr ago. 

\item The close agreement between the sulfur abundances in Orion and the Sun provides 
indirect support to previous claims by \citet{lan08} and \citet{chl06}
that the young Orion stars can be used as good proxies in order to ascertain 
the solar abundances of noble gases, such as Ne and Ar, whose abundances are not measurable in
the solar photosphere. 

\item The abundances of main-sequence B stars overlap with the gas-phase abundance 
derived by \citet{est04} for the Orion Nebula.
The similarity between these abundances indicates that sulfur is undepleted in the Orion Nebula.
Contrary to elements such as C and Fe, for which important depletion exists, 
sulfur appears to be a remarkably direct metallicity indicator of the Galactic disk 
(except perhaps in very dense regions).

\item The sulfur abundance results for Orion are added to a database of abundances previously
published for OB main sequence stars along the Galactic Disk.
The new sulfur abundance gradient computed for a sample of 50 OB stars 
is $A(S)=(7.481\pm 0.097) -(0.037\pm 0.012) R_g$. 
The sulfur gradient is very similar to the flat gradient of $-0.031 \pm 0.012$ dex~Kpc$^{-1}$ 
previously obtained for oxygen in \citet{dec04} and in agreement with  standard chemical 
evolution model predictions such as \citet{cmr01} and \citet{ces07}.

\end{enumerate}

\acknowledgments
We thank Keith Butler and Ivan Hubeny for discussions. We thank the referee for good suggestions.

%%TABLES

\clearpage
\input{tab1}

\input{tab2}

\input{tab3}

\clearpage

%%FIGURES

\clearpage

\clearpage

\clearpage

\clearpage

\clearpage

\end{document}

%% file: tab1.tex
\begin{deluxetable}{cccccccc}
\tablecaption{Stellar Parameters  and Sulfur Abundances \label{tbl-1}}
\tablewidth{0pt}
\tablehead{
\colhead{Star} &
\colhead{$T_{{\rm eff}}$} & 
\colhead{$\log g $} & 
\colhead{$\xi$} &
\colhead{$V \sin i$} &
\colhead{ A(S {\sc ii})} &
\colhead{ A(S {\sc iii})} & 
\colhead{ A(S)} \\
\colhead{} &
\colhead{(K)} & 
\colhead{} & 
\colhead{(km s$^{-1}$)} &
\colhead{(km s$^{-1}$)} &
\colhead{} &
\colhead{} &
\colhead{} }
\startdata
HD 35039 &  20550  & 3.74  & 8.0 & 9$\pm$ 1 & 7.18$\pm$0.16[16] & 7.14$\pm$0.13[3] & 7.18$\pm$0.15 [19] \\
HD 35299 &  24000  & 4.25  & 8.0 & 12$\pm$2 & 7.12$\pm$0.13[16] & 7.06$\pm$0.07[3] & 7.11$\pm$0.13 [19]\\
HD 35912 &  19590  & 4.20  & 8.0 & 14$\pm$2 & 7.14$\pm$0.14[15] & 7.15$\pm$0.02[2] & 7.14$\pm$0.13 [17] \\
HD 36285 &  21930  & 4.40  & 8.0 & 14$\pm$2 & 7.15$\pm$0.13[15] & 7.17$\pm$0.10[2] & 7.15$\pm$0.13 [17] \\
HD 36351*&  22520  & 4.23  & 9.0 & 31$\pm$4 & 7.09$\pm$0.10[7]  & 7.15$\pm$0.01[2] & 7.11$\pm$0.09 [9]\\
HD 36591*&  26860  & 4.26  & 9.0 & 11$\pm$2 & 7.06$\pm$0.03[2]  & 7.09$\pm$0.11[2] & 7.07$\pm$0.07 [4]\\
HD 36959 &  24890  & 4.41  & 6.0 & 13$\pm$3 & 7.13$\pm$0.14[9]  & 7.24$\pm$0.14[3] & 7.16$\pm$0.14 [12]\\
HD 37209 &  24050  & 4.13  &10.0 & 44$\pm$8 & 7.22$\pm$0.05[5]  & 7.20$\pm$0.12[2] & 7.22$\pm$0.06 [7]\\
HD 37356 &  22370  & 4.13  & 9.0 & 20$\pm$3 & 7.15$\pm$0.14[13] & 7.05$\pm$0.14[3] & 7.13$\pm$0.14 [16]\\
HD 37744*&  24000  & 4.35  & 7.0 & 38$\pm$4 & 7.27$\pm$0.14[3]  & 7.20$\pm$0.04[2] & 7.24$\pm$0.11 [5]\\
\enddata
\tablenotetext{*}{Revised T$_{\rm eff}$ and $\log g$}
\end{deluxetable}

%% file: tab2.tex
\begin{deluxetable}{cccr}
%\tabletypesize{\scriptsize}
\tablecaption{Atomic Data \label{tbl-2}}
\tablewidth{0pt}
\tablehead{
\colhead{Species} & 
\colhead{Wavelength} &
\colhead{$\chi$} & 
\colhead{$\log gf $}  \\
\colhead{} & 
\colhead{(\AA)} &
\colhead{(eV)} & 
\colhead{} }
\startdata
S{\sc ii}  & 4162.66 &  15.944  &    0.789 \\
           & 4217.18 &  15.944  & $-$0.145 \\
           & 4269.72 &  16.092  & $-$0.112 \\  
           & 4278.51 &  16.092  & $-$0.111 \\ 
           & 4294.40 &  16.135  &    0.568 \\
           & 4463.58 &  15.944  &    0.213 \\
           & 4483.43 &  15.899  & $-$0.064 \\
           & 4486.63 &  15.867  & $-$0.463 \\
           & 4524.72 &  15.068  &    0.031 \\
           & 4656.76 &  13.584  & $-$0.519 \\
           & 4729.44 &  16.100  & $-$0.139 \\
           & 4815.55 &  13.672  & $-$0.051 \\
           & 4824.07 &  16.265  &    0.062 \\
           & 5014.07 &  14.068  &    0.054 \\
           & 4991.97 &  13.617  & $-$0.739 \\
           & 5032.43 &  13.672  &    0.157 \\
S{\sc iii} & 4284.88 &  18.193  & $-$0.277 \\
           & 4361.48 &  18.244  & $-$0.760 \\
           & 4364.68 &  18.318  & $-$0.846 \\
\enddata
\end{deluxetable}

%% file: tab3.tex
\begin{deluxetable}{ccc}
%\tabletypesize{\scriptsize}
\tablecaption{Abundance Uncertainties \label{tbl-3}}
\tablewidth{0pt}
\tablehead{
\colhead{ } & 
\colhead{S {\sc ii}} &
\colhead{S {\sc iii}} }
\startdata
$\delta (T_{eff})=+4$\%   &   +0.09  &  $-0.08$  \\
$\delta (\log g)=+0.1 $   &   +0.02  &  +0.01  \\
$\delta (\xi)=+1.5$      &  $-0.04$   & $-0.04$ \\  
$\delta (gf)=+10$\%       &   +0.04  &  +0.04   \\ 
$\delta {\rm (continuum)}=+0.5$\%   &  +0.02  &   +0.03\\
$\delta {\rm (total)}$   &  +0.11 &   +0.10 \\
\enddata
\end{deluxetable}